\documentclass{article}

\usepackage[final]{neurips_2024}

\usepackage[utf8]{inputenc} 
\usepackage[T1]{fontenc}    
\usepackage{hyperref}       
\usepackage{url}            
\usepackage{booktabs}       
\usepackage{amsfonts}       
\usepackage{nicefrac}       
\usepackage{microtype}      
\usepackage{xcolor}         
\usepackage{graphicx}
\usepackage{amsmath}
\usepackage{natbib}
\usepackage{graphicx} 
\usepackage{subcaption} 
\usepackage{wrapfig}
\usepackage{multicol}
\usepackage{multirow}
\usepackage{pifont}
\usepackage{algorithm}
\usepackage{algpseudocode}
\usepackage{array}
\usepackage{authblk}
\usepackage{fancyhdr}

\setcitestyle{square,numbers}
\title{Mortality Prediction of Pulmonary Embolism Patients with Deep Learning and XGBoost}

\author[1]{Yalcin Tur}
\author[2]{Vedat Cicek}
\author[3]{Tufan Cinar}
\author[2]{Elif Keles}
\author[2]{Bradlay D. Allen}
\author[2]{Hatice Savas}
\author[2]{Gorkem Durak}
\author[2]{Alpay Medetalibeyoglu}
\author[2]{Ulas Bagci$^*$}
\affil[1]{Department of Computer Science, Stanford University, Stanford, CA, USA.}
\affil[2]{Department of Radiology, Northwestern University, Chicago, IL, USA.} 
\affil[3]{School of Medicine, Univesity of Maryland, Baltimore, MD, USA.}

\begin{document}

\maketitle
\def\thefootnote{*}\footnotetext{Corresponding author: \texttt{ulas.bagci@northwestern.edu}}

\begin{abstract}
Pulmonary Embolism (PE) is a serious cardiovascular condition that remains a leading cause of mortality and critical illness, underscoring the need for enhanced diagnostic strategies. Conventional clinical methods have limited success in predicting 30-day in-hospital mortality of PE patients.  In this study, we present a new algorithm, called  PEP-Net, for 30-day mortality prediction of PE patients based on the initial imaging data (CT) that opportunistically integrates a 3D Residual Network (3DResNet) with Extreme Gradient Boosting (XGBoost) algorithm with patient level binary labels without annotations of the emboli and its extent. Our proposed system offers a comprehensive prediction strategy by handling class imbalance problems, reducing overfitting via regularization, and reducing the prediction variance for more stable predictions. PEP-Net was tested in a cohort of 193 volumetric CT  scans diagnosed with Acute PE, and it demonstrated a superior performance by significantly outperforming baseline models (76-78\%) with an accuracy of 94.5\% (±0.3) and 94.0\% (±0.7) when the input image is either lung region (Lung-ROI)  or heart region  (Cardiac-ROI). Our results advance PE prognostics by using only initial imaging data, setting a new benchmark in the field. While purely deep learning models have become the go-to for many medical classification (diagnostic) tasks, combined ResNet and XGBoost models herein outperform sole deep learning models due to a potential reason for having lack of enough data.  
\end{abstract}

\section{Introduction}

\begin{figure}[h]
   \centering
  \includegraphics[width=0.9\textwidth]{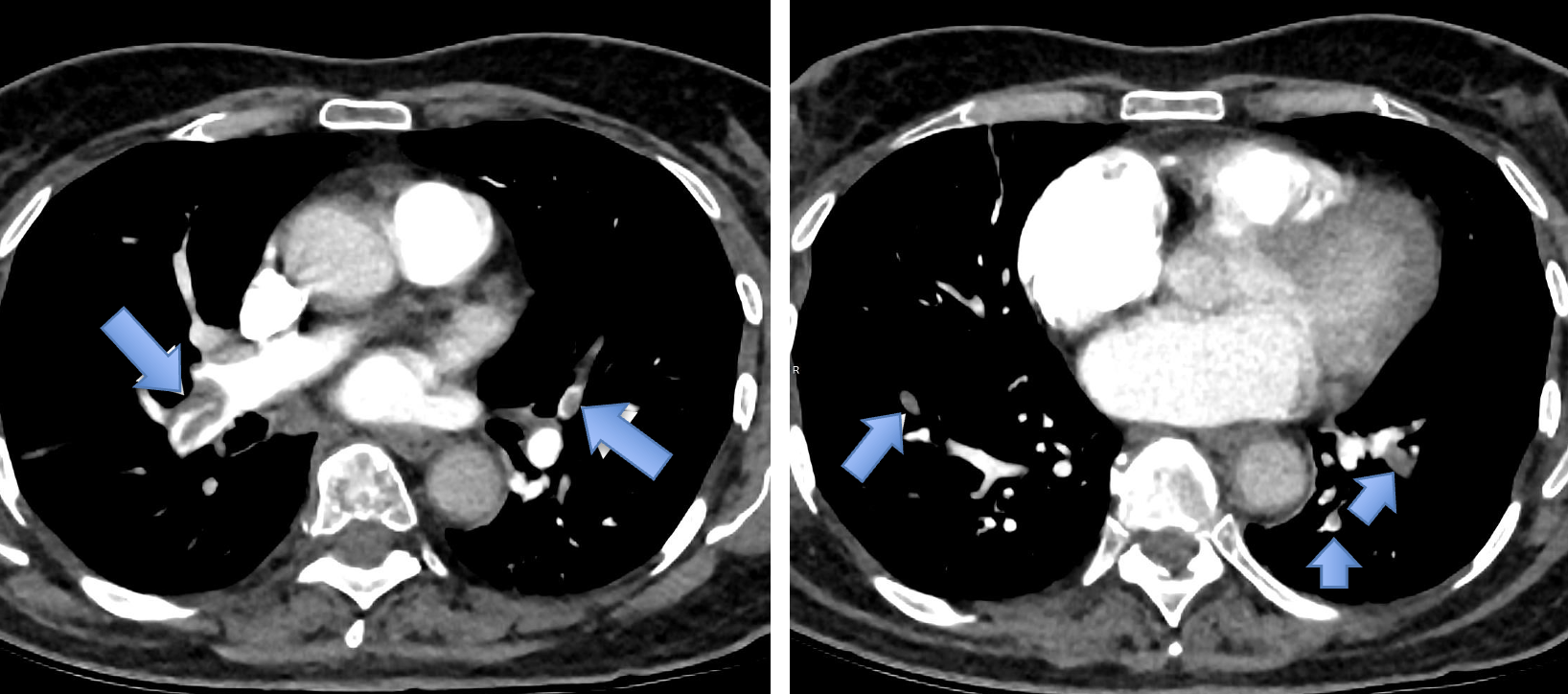}
    \caption{We have lobar (left) and distal PE (right) cases are illustrated, respectively.}
    \label{fig:pe}
\end{figure}

Pulmonary embolism (PE) is a severe and sometimes fatal condition caused by a blockage in one or more lung arteries, usually due to a blood clot that forms in the veins of the legs or pelvis~\cite{georgilis2023understanding}. This blockage can disrupt blood flow and increase pressure in the heart's right ventricle. PE is a leading cause of death in hospitals, ranking just behind heart attacks and strokes. Studies indicate that every year, between 39 and 115 out of every 100,000 people experience PE~\cite{dentali2016time}. 

In the United States, PE is believed to result in around 300,000 deaths each year, costing the healthcare system between 7 and 10 billion dollars annually. For people aged 65 and older, the in-hospital death rate from PE is about 4\%, with the six-month mortality rate reaching 20\%. Additionally, 15\% of patients in the hospital for PE are readmitted within the first 30 days~\cite{wendelboe2016global}. The symptoms of PE can vary widely, from no symptoms to severe cases leading to shock. Diagnosing PE correctly and quickly is critically important for two main reasons. First, the primary treatment for PE, which involves thinning the blood to prevent clots, is within the risk of bleeding. Inaccurate diagnosis could lead to death from either the PE itself or from dangerous bleeding complications that are preventable. Thus, a swift and precise diagnosis and particularly prognosis (\textit{i.e., predicting patient outcome}) is crucial in effectively managing patients. 

\textbf{Current guidelines for diagnosis.} Diagnosing (PE) involves a combination of clinical assessment, imaging studies (CT), and laboratory tests, aimed at confirming the presence of a blood clot in the pulmonary arteries. The process is mostly guided by the probability of PE based on initial clinical evaluation, including symptoms and risk factors.
The clinical symptoms of PE are not sensitive or specific enough to definitively diagnose or rule out the condition. For this reason, many scoring systems have been developed to guide clinicians in considering or excluding PE in the differential diagnosis. The 2019 European Society of Cardiology (ESC) Acute Pulmonary Embolism guidelines recommend using the Geneva and WELLS scores as diagnostic aids~\cite{konstantinides20202019}. In clinical practice, the commonly used WELLS score calculates the probability of PE~\cite{aydogdu2014wells}. Regardless of the score, patients with confirmed PE are likely to fall into three categories 10\% for low probability, 30\% and  for moderate probability, and 65\% for high probability~\cite{ceriani2010clinical}. Given PE's clinical significance and the current low diagnostic accuracy rates, there is a pressing need to develop new diagnostic approaches. In this study, we focus on prognosis of PE patients, which is rather a less frequently studied important topic, and the patients involved in our study were already diagnosed with PE when admitted to hospital, separate from control group.

\begin{figure}
    \centering
    \includegraphics[width=0.95\textwidth]{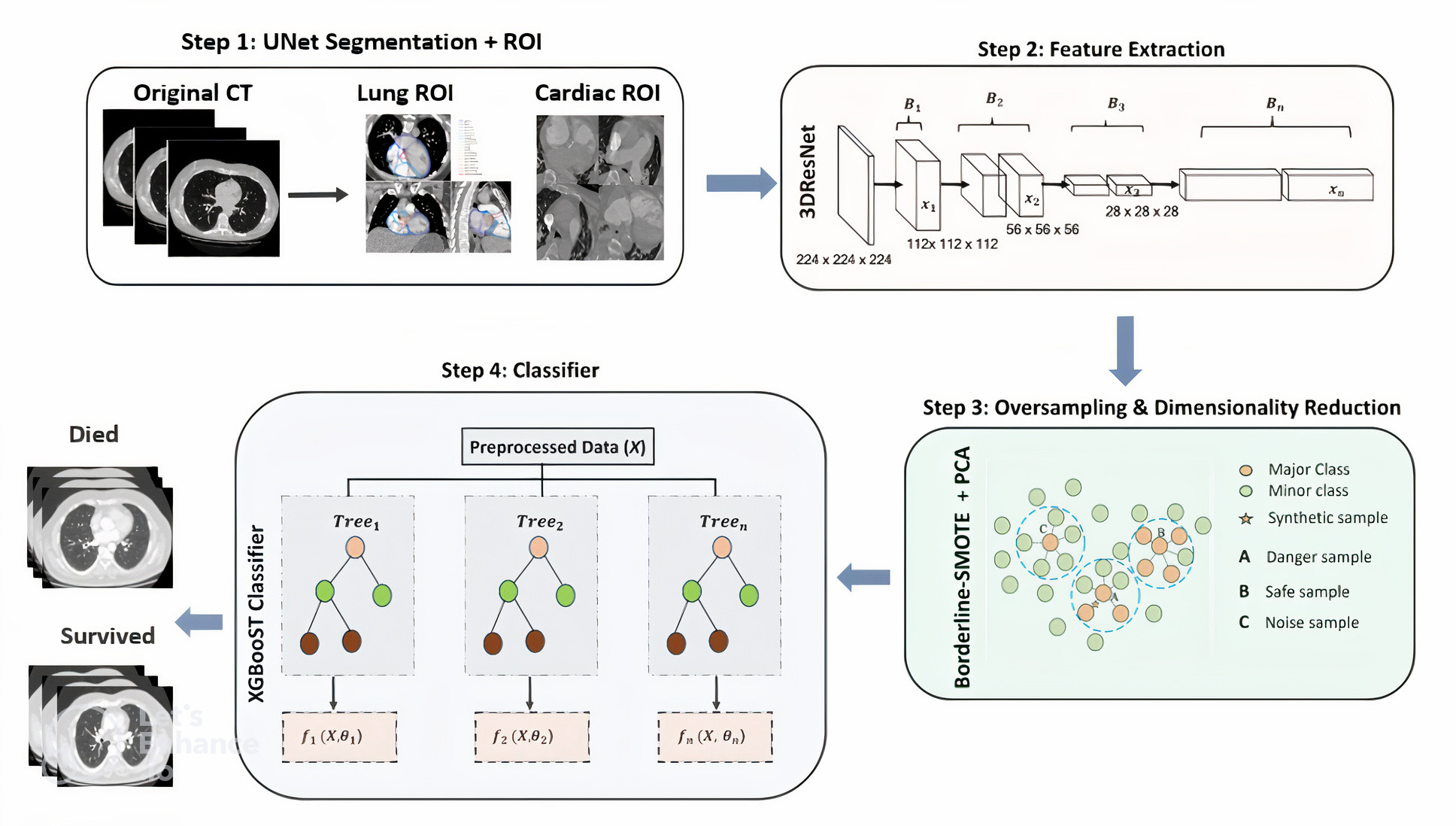}
    \caption{Proposed PEP-Net architecture for mortality prediction for PE patients include four consecutive steps. Lung and cardiac region localizations (rough ROIs), feature extraction with deep nets, oversampling and dimensionality reduction, and an optimized classifier.}
    \label{fig:ppnet}
\end{figure}

\textbf{Current guidelines for prognosis.} The prognosis of PE patients is usually determined by hemodynamics. Large and multiple emboli can suddenly raise pulmonary vascular resistance and lead to an increased afterload that the right ventricle (RV) cannot tolerate, resulting in sudden death. Goldhaber et al. (1999) found that the 14-day mortality rate for PE was over 20\%~\cite{goldhaber1999acute}, and a study conducted by Bikdeli et al. (2022), found that the cumulative overall mortality rate at three months was 8.65\%~\cite{bikdeli2022clinical}. 
The imaging findings from a CT (computed tomography) scan can provide prognostic information in addition to diagnosis. For example, the presence of a saddle embolus (an embolus straddling the bifurcation of the main pulmonary artery) is associated with a higher mortality rate. \textbf{F}\textbf{igure 1} shows two example PE cases (lobar and distal). Right ventricular strain, clot burden, pulmonary artery obstruction, and infarction  are some of the imaging findings associated with prognosis of PE patients.

\textbf{Critical need for new methods for prognosis.} The symptoms of PE can vary widely, from no symptoms to severe cases leading to shock. PE diagnosis and prognosis are critically important for two main reasons. First, the primary treatment for PE, which involves thinning the blood to prevent clots, carries a significant risk of bleeding. A wrong diagnosis could lead to unnecessary death from either the PE itself or from dangerous bleeding complications. Thus, a swift prognosis after a precise diagnosis (\textit{i.e., predicting patient outcome}) is crucial in effective management of patients. Unfortunately, current clinical standards for prognosis falls short and there is a significant need for a better outcome prediction for patients with PE ~\cite{douma2010acute}.




\textbf{PESI score is promising but not definitive enough.} Distinguishing between low-risk and high-risk patients in PE is important for determining the treatment approach and supporting the prognosis. For this purpose, several prognostic classifications have been developed for PE patients. The most commonly used prognostic method is the PESI scoring, designed by Aujesky and colleagues in 2005, which is recommended for use in the ESC 2019 PE guidelines. The PESI score includes 11 clinical variables, including demographic characteristics such as the presence of heart failure and a history of cancer, as well as vital signs such as hypotension and tachycardia~\cite{aujesky2005derivation}. It predicts the 30-day mortality of PE patients with a sensitivity of 85-90\% and a specificity of 35-45\%~\cite{zhou2012prognostic}. 

\textbf{Summary of our contributions:} Despite the high mortality rates, with a successful prognostic classification, PE remains one of the leading causes of preventable in-hospital deaths. Therefore, the development of new prognostic predictors can enable patients to receive appropriate treatment early, leading to a decrease in possible mortality and morbidity. Our study in this paper belongs to this class of research. Our contributions can be summarized as follows:
\begin{enumerate}
    \item 30-day mortality predictions using deep learning based approaches has not been explored thoroughly, which is one of the main motivations behind our study herein. We introduced a novel deep learning-based approach called PEP-Net (pulmonary embolism prognosis network), which includes several interconnected modules, carefully designed to automatically analyze CT scans. 
    \item  These modules includes U-Net for "roughly" estimate the location of lungs and heart,  3D-Residual Network (3DResNet) for extracting volumetric hierarchical features, principal component analysis (PCA) for feature dimensionality reduction, BorderlineSMOTE (B-SMOTE) for solving class-imbalance problem, and XGBoost to take learned features and focus on the relationship that are most predictive of the outcome. 
    \item We have devised U-Net segmentor inside the PEP-Net to have lung and cardiac regions to be determined automatically and used as input to the predictor for 30-day mortality prediction. Separating cardiac and lung regions is aimed for combining findings from lobar and distal cases in case needed. 
    \item  Our comprehensive results demonstrate that the PEP-Net outperforms other baselines, achieving superior performance (AUC=0.92) while the closest results obtained with ResNet18 architecture was of AUC=0.73. 
\end{enumerate}



\section{Methods}
\textbf{Data:} In this retrospective cohort study (IRB approved), we evaluated a total of 193 computed tomography (CT) scans obtained from Sultan Abdulhamid Han Education and Research Hospital, Istanbul, Turkey. Patients were presented with clinical symptoms suggestive of PE, and their 30-day  mortality predictions were recorded. CT scans were performed using a Canon Aquilion 128 scanner CT when they were admitted to the hospital. The scanning protocol was standardized to acquire images with a slice thickness of 1 mm. The scanning protocol included the following parameters: tube voltage, 120 kV; tube current time product, 150-300 mAs; pitch, 0.75-0.85. The in-plane spatial resolution of the images was further enhanced with a pixel matrix of 512x512 and a pixel spacing of 1 mm x 1 mm in plane resolution. To optimize contrast resolution, a non-ionic iodinated contrast agent was administered intravenously (i.e., 1 cc of intravenous contrast agent per kilogram). To maintain the privacy and confidentiality of patients, all scans were anonymized. 

Table~\ref{tab:data-distribution} summarizes patient cohort categorization based on their status within 30 days following the CT scan. Out of the 193 patients, 38 (19.69\%) died within 30 days of the scan, while 155 (80.31\%) survived beyond this period. This imbalanced data cohort includes only patient-level label for mortality. 

\begin{table}[ht]
\centering
\caption{30-day mortality information of PE patients.}
\label{tab:data-distribution}
\begin{tabular}{lcc}
\hline
\textbf{Patient Status} & \textbf{Number of Patients} & \textbf{Percentage} \\ \hline
Died within 30 days & 38 & 19.69\% \\
Survived beyond 30 days & 155 & 80.31\% \\ \hline
\textbf{Total} & 193 & 100\% \\ \hline
\end{tabular}%
\end{table}

\textbf{Deep Learning Prognosis:} Our proposed approach consists of several steps: identifying the Lung-ROI and Cardiac-ROI, adopting a 3DResNet~\cite{zhou2012prognostic} model for feature extraction, using PCA for dimensionality reduction, applying B-SMOTE to handle class imbalance problems, and finally employing a fine-tuned, optimized XGBoost classifier to train and evaluate the model's performance as depicted in \textbf{Figure~\ref{fig:ppnet}}.

\textbf{Determining Lung-ROI and Cardiac-ROI:} Identifying both the Lung-ROI and Cardiac-ROI requires a segmentation using a conventional 3D U-Net model to create 'rough' segmentation masks~\cite{demir2021information,mansoor,caban}. Subsequently, we use these masks to select the ROI encapsulating lung regions by identifying non-zero boundary voxels from the segmentation images, defining a tight bounding box. A similar approach was applied to identify the Cardiac regions~\cite{mortazi2017cardiacnet}. We used publicly available pre-trained models and checkpoints for lung and heart segmentation from \textit{TotalSegmentator}~\cite{wasserthal2023totalsegmentator}.

\textbf{3DResNet with a new convolutional layer:}
Our study employs a pre-trained 3DResNet model, specifically the ResNet18 variant with 18 layers, known for its fine-tuned weights acquired from substantial training data. These pre-trained weights enhance performance, particularly with smaller datasets, by expediting convergence. The model structure, adapted from the original pre-trained model, follows a sequential design but omits the first and last layers. The first layer, initially intended for 3-channel RGB images, is replaced to accommodate grayscale CT images with a single channel. Likewise, the final fully connected layer, designed for classification, is removed to focus on feature extraction.

We also introduce a new 3D convolutional layer to process the 1-channel input, aligning with grayscale CT images and generating a 64-channel feature map. This layer maintains the original first layer's kernel size, stride, and padding parameters. Its weights are trained from scratch during training, with a kernel size of (7, 7, 7), a stride of (2, 2, 2), and a padding of (3, 3, 3) to preserve spatial dimensions post-convolution by adding border pixels. (\textbf{Figure 2}, step 2).


\textbf{Oversampling and Feature Selection:} An advanced oversampling technique, B-SMOTE, is applied in our work because we have a class imbalance problem (20\% vs 80\% distribution of patients with PE) and classifier performance is suboptimal with class imbalance. We apply default parameters, including a sampling strategy of 'auto', a k-neighbors value of 5, and an m-neighbors value of 10. For a given minority class sample  $x_i$ and its selected neighbor $x_{in}$, a synthetic sample $x_{new}$ is generated as follows:
\begin{equation}
x_{new} = x_i + \lambda \cdot (x_{in} - x_i),
\end{equation}
\noindent where \( \lambda \) is a random number between 0 and 1. 

Next, we apply PCA (principal component analysis) based dimensionality reduction (feature selection) to our system. PCA is widely used for feature selection, especially after a large number of features have been generated by deep learning networks. Determining the number of principal components for PCA was based on experimentation, with a range of 50 to 150 components tested. The optimal performance is achieved with 100 components (empirically determined).

\textbf{Classification with XGBoost:} The XGBoost classifier is used with the objective set as `\textit{binary:logistic}', indicating a binary classification problem with logistic regression for probability prediction. Key parameters include a learning rate of 0.1, which balances model training speed and performance, and a maximum tree depth of 3 to limit model complexity and mitigate overfitting. The model is trained with 100 boosting rounds. The \textit{scale\_pos\_weight }parameter, calculated as the ratio of the number of negative cases to the number of positive cases in the training data, is used to give higher importance to the minority class during training. For a given iteration \( i \), the update equation for the XGBoost classifier \( F_i(x) \) can be expressed as follows:
\begin{equation}
F_i(x) = F_{i-1}(x) + \eta \cdot h_i(x),
\end{equation}
\noindent where \( F_{i-1}(x) \) is the model at iteration \( i-1 \), \( \eta \) is the learning rate, and \( h_i(x) \) is the new decision tree added at iteration \( i \).








\textbf{Further Details on Training:} PEP-Net and other baselines undergo training for up to 100 epochs, utilizing \textit{Sparse Categorical Cross Entropy} as the loss function, with ADAM optimizer set along with a learning rate of 0.0001. The training is carried out on a high-performance GPU server equipped with 1 NVIDIA A100 GPU and 80GB memory, operating under Debian Linux. We evaluate the performance of each method based on metrics such as Accuracy, Area under the Curve (AUC), Sensitivity, and Specificity. We have used five-fold cross-validation during the experiments, ensuring robustness and reliability in our results. 


\section{Results}
We compared the performance of our proposed PEP-Net method with three deep network models: ResNet18~\cite{zhou2012prognostic}, ResNet50~\cite{xue2023region}, and EfficientNetB0~\cite{zheng2023modified}, respectively. Our objective was to assess the efficacy of these models in predicting 30-day mortality in PE patients using either Lung-ROI or Cardiac-ROI inputs.


Table~\ref{tab:tabres} compares the performance of different neural network architectures on a classification task using various metrics. ResNet18 shows a decent, but not the best among the methods listed. Its AUC (0.581) is relatively low, indicating that the model is not as good at distinguishing between the classes. The sensitivity, which measures the true positive rate  (0.782), suggesting that it can identify the positive cases fairly well. However, its specificity of 0.535 is low, indicating it's less effective at identifying the negative cases. 
ResNet50 shows improved accuracy over ResNet18, suggesting it's better at classifying the correct outcomes overall. The AUC is slightly better than ResNet18, but still indicates room for improvement in model discrimination. EfficientNetB0 has a lower accuracy than both ResNet models, but it surpasses them with the highest AUC, indicating better performance in distinguishing between the positive and negative classes. Both sensitivity and specificity metrics showing that EfficientNetB0 has a balanced performance in identifying positive and negative cases respectively.

The proposed PEP-Net significantly outperforms the other models across all metrics. It has the highest accuracy, suggesting it's highly effective at making correct predictions. The AUC (0.91) is excellent, indicating a superior ability to distinguish between the classes. These outstanding results underscore the potential of our proposed model for improving mortality prediction in PE patients.

\begin{table}[h]
\caption{Performance comparison of PEP-Net and baselines for 30-day mortality prediction. {\color{blue}Input: Lung-ROI.}}
\label{tab:tabres}
\begin{tabular}{lcccc}
\hline
\textbf{Method} & \textbf{Accuracy} & \textbf{AUC} & \textbf{Sensitivity} & \textbf{Specificity} \\ \hline
ResNet18 & \(0.783 \pm 0.029\) & \(0.581 \pm 0.101\) & \(0.782 \pm 0.184\) & \(0.535 \pm 0.198\) \\ 
ResNet50 & \(0.803 \pm 0.010\) & \(0.583 \pm 0.110\) & \(0.525 \pm 0.374\) & \(0.703 \pm 0.288\) \\ 
EfficientNetB0 & \(0.716 \pm 0.127\) & \(0.670 \pm 0.073\) & \(0.661 \pm 0.187\) & \(0.774 \pm 0.167\) \\ \hline
\textbf{PEP-Net} & \textbf{0.945 \(\pm\) 0.003} & \textbf{0.917 \(\pm\) 0.007} & \textbf{0.977 \(\pm\) 0.002} & \textbf{0.874 \(\pm\) 0.013} \\ \hline
\end{tabular}%
\end{table}

\begin{figure*}
    \includegraphics[width=.4\textwidth]{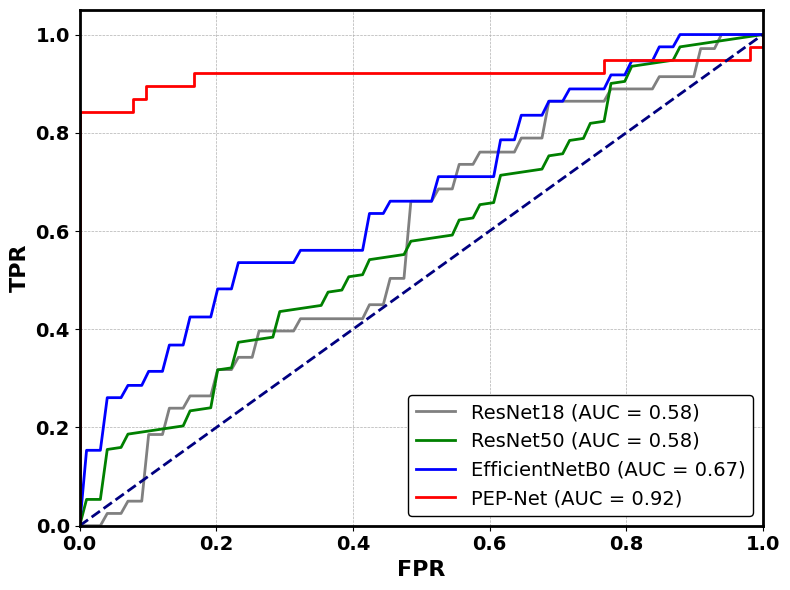}
     \includegraphics[width=.6\textwidth]{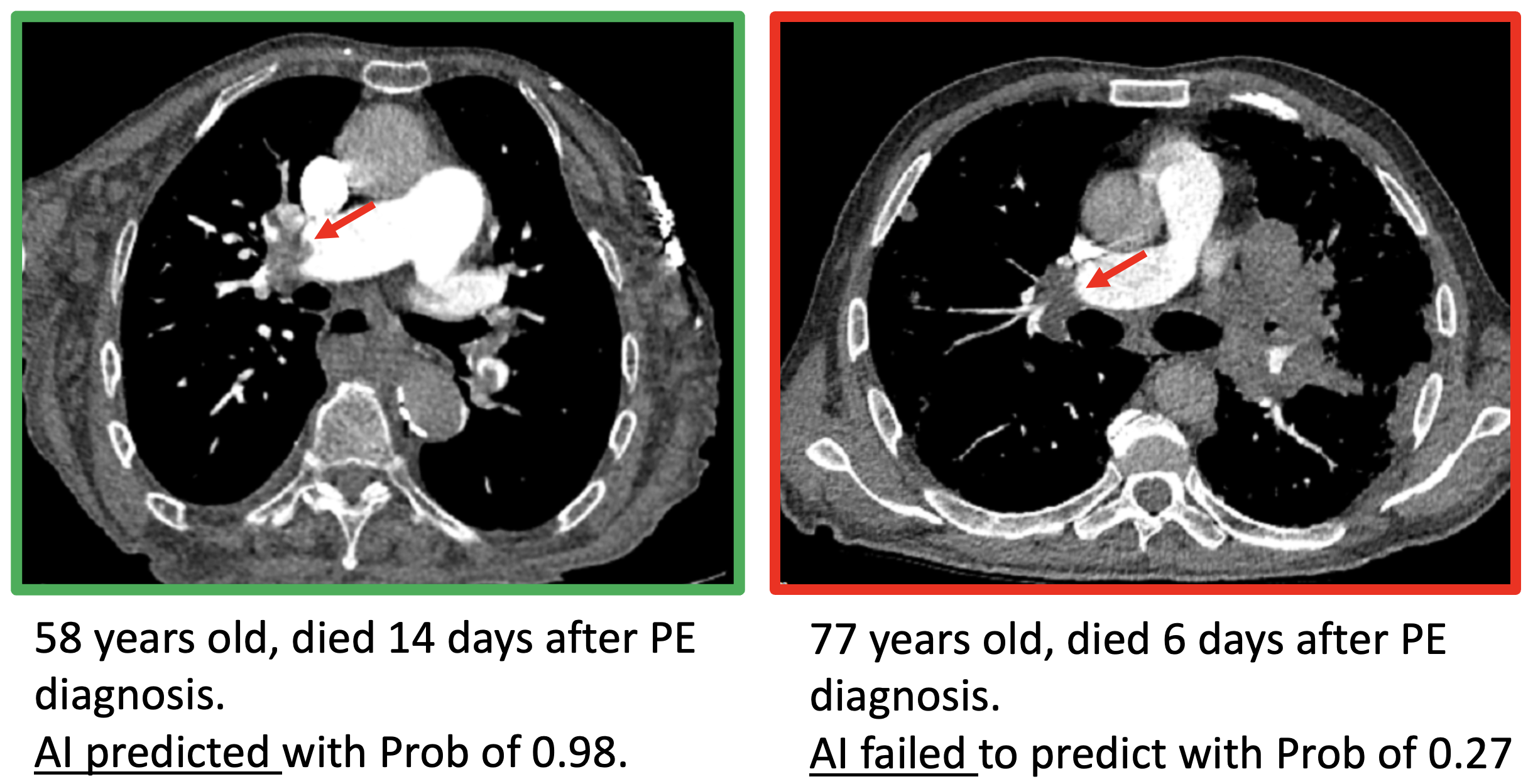}
    \caption{\textbf{Left:} ROC-AUC curve of different DL-based approaches used in this study; TPR-- True Positive Rate, FPR-- False Positive Rate. \textbf{Middle:} An example CT scan that PEP-Net predicts the 30-day mortality of PE patient with high probability. \textbf{Right:} An example CT scan that PEP-Net fails to predict the patient's 30-day mortality. Very subtle change is observed.}
    \label{fig:roc-auc}

\end{figure*}

Table~\ref{tab:heartres} presents a performance comparison of PEP-Net and other referenced CNN models for 30-day mortality prediction but this time our input to the predictor is Cardiac-ROI. We observed that our PEP-Net outperforms the referenced models similar to that of with Lung-ROI, with an accuracy of 0.940 (± 0.007) and an AUC of 0.901. While Cardiac-ROI results are promising, they have some interesting implications too. For example, Cardiac-ROI based results are very close to Lung-ROI based results, indicating that the algorithm does not only look for emboli regions and finding some changes in the cardiac or vessels that is not visible to human eye at first instance. In our future work, we will develop a precise visual explanation algorithm to identify distal PE's effect on the cardiac images. 

\begin{table}[h]
\caption{Performance comparison of PEP-Net and baselines for 30-day mortality prediction. {\color{blue}Input: Cardiac-ROI.}}
\label{tab:heartres}

\begin{tabular}{lcccc}
\hline
\textbf{Method} & \textbf{Accuracy} & \textbf{AUC} & \textbf{Sensitivity} & \textbf{Specificity} \\ \hline
ResNet18 & \(0.693 \pm 0.159\) & \(0.733 \pm 0.100\) & \(0.889 \pm 0.106\) & \(0.582 \pm 0.174\) \\ 
ResNet50 & \(0.709 \pm 0.176\) & \(0.563 \pm 0.148\) & \(0.607 \pm 0.334\) & \(0.654 \pm 0.294\) \\
EfficientNetB0 & \(0.765 \pm 0.141\) & \(0.660 \pm 0.131\) & \(0.746 \pm 0.174\) & \(0.721 \pm 0.207\) \\ \hline
\textbf{PEP-Net} & \textbf{0.940 \(\pm\) 0.007} & \textbf{0.901 \(\pm\) 0.008} & \textbf{0.962 \(\pm\) 0.004}& \textbf{0.852 \(\pm\) 0.013} \\ \hline
\end{tabular}%
\end{table}

Figure~\ref{fig:roc-auc} illustrates the superior performance of PEP-Net as compared to other CNN models. As shown, AUC score for PEP-Net with Lung-ROI is significantly higher than that of the referenced CNN models. Overall, our algorithm \textit{misclassified} only 7 patients. For the three patients (alive), our model made the wrong prediction (death). Of these patients, one had the lobar type,  and two had the distal branch type. Also, for the four patients (died),  our model made the wrong prediction (alive). Of these patients, two had the main branch type, one had the lobar type, and one had the distal branch type. Figure 3 (right) shows one of the misclassified patients The same figure middle illustrates a case where PEP-Net predicted the prognosis with high probability although emboli is subtle (red arrow).


\section{Discussion and Concluding Remarks}
We introduce a new deep learning model, PEP-Net, tailored for the prognostication of Pulmonary Embolism (PE), a critical cardiovascular ailment with high mortality rates. Several factors contributes to the success of our model in this context. Firstly, we employed BorderlineSMOTE to address the dataset's imbalance, generating synthetic samples for the minority class. This balancing technique proved effective in improving model generalization. Additionally, we incorporated PCA to reduce dimensionality, enhancing feature representation. This integration, alongside the 3DResNet architecture and XGBoost algorithm, created a powerful combination. ResNet excels at feature extraction and learning representations from images, while XGBoost, renowned for its ability to handle imbalanced datasets, plays a pivotal role. XGBoost, in combination with BorderlineSMOTE, focuses on the minority class, effectively addressing data imbalance by giving greater attention to challenging instances. This boosting approach allows our model to learn from its mistakes and refine its predictions, which is particularly vital in imbalanced medical datasets. Validated on 193 CT scans, PEP-Net achieved remarkable accuracy rates above 94\%.

While traditional prognostic models fall short, PEP-Net provides state-of-the-art results by integrating 3DResNet and XGBoost to masterfully address class imbalances and enhance prediction stability. Validated on 193 CT scans, PEP-Net achieved remarkable accuracy rates above 94\%. Notably, the model's specificity improved when analyzing Lung-ROI compared to Cardiac-ROI, suggesting the Lung region's significant role in identifying non-risk cases especially when the embolism is distal. However, it should be also noted the difference is small, potentially suggesting that there maybe some changes in the cardiac anatomy not quite visible to human-eye, and still suggesting PE formation. Compared to PEP-Net's performance, referenced CNN models, including ResNet18 and ResNet50, showed only marginal performance gains in Cardiac-ROI over Lung-ROI, yet none matched the PEP-Net's robust metrics. 


While our study shows promising results, there are some limitations which we aim to address in the future. First, we will repeat our study in a multi-center manner to more precisely assesses the generalization of the proposed algorithm. Getting data from multiple center is a challenge at the moment. Next, we did not incorporate clinical parameters, which could provide valuable insights for developing a more robust prediction model. Our study can also benefit from a more advanced backbone models such as ViT but one needs to be aware of data-hungry nature of Transformers while the data is limited in our case.

One may also ask if solely designed deep networks (ViT, Mamba, and others) can solve this problem without the need for XGBoost or other separate classifiers. The decision to combine traditional machine learning techniques, such as XGBoost, with deep learning architectures, like 3DResNet, for classification tasks is a strategic approach that leverages the strengths of both methodologies to achieve superior performance, not only the lack of enough data to attend. We would like to note that this choice does not inherently imply that fully deep learning methods are incapable of capturing good results; rather, it highlights a nuanced understanding of the problem space and the tools available. One immediate benefit by combining these two methods was also to reduce training time and computational resource requirements, making it a practical solution for many real-world applications.

\bibliography{strings,refs}

\end{document}